\documentclass[11pt,a4paper]{article}
\usepackage{jheppub}

\usepackage{epsfig}
\usepackage{cancel}

\usepackage{amsmath, amsthm, amsfonts, amssymb, latexsym}
\usepackage{upgreek}
\usepackage[caption=false]{subfig}

\usepackage{url}
\usepackage{ifpdf}

\def\beq{\begin{equation}}
\def\eeq{\end{equation}}

\def\bequ{\begin{equation}}
\def\eequ{\end{equation}}

\title{\Large \textsc{ScannerS}: Constraining the phase diagram of a complex scalar singlet at the LHC}

\author[a]{Rita Coimbra,} 
\emailAdd{rita.coimbra@coimbra.lip.pt}
\author[b]{Marco O. P. Sampaio,} 
\emailAdd{msampaio@ua.pt}
\author[c]{Rui Santos}
\emailAdd{rsantos@cii.fc.ul.pt}
\affiliation[a]{ LIP, Departamento de F\'{\i}sica,  Universidade de Coimbra\\
 3004-516 Coimbra, Portugal} 
\affiliation[b]{Departamento de F\'\i sica da Universidade de Aveiro and I3N \\ 
Campus de Santiago, 3810-183 Aveiro, Portugal} 
\affiliation[c]{
 Instituto Superior de Engenharia de Lisboa - ISEL \\
 1959-007 Lisboa, Portugal and \\
 Centro de F\'\i sica Te\'orica e Computacional,
Universidade de Lisboa  \\
1649-003 Lisboa, Portugal
}

\keywords{Higgs boson, Beyond Standard Model}

\abstract{We present the first version of a new tool to scan the parameter space of generic scalar potentials, \textsc{ScannerS}~\cite{ScannerS}.  
The main goal of \textsc{ScannerS} is to help distinguish between different patterns of symmetry breaking for each scalar potential.
In this work we use it to investigate the possibility of excluding regions of the phase diagram 
of several versions of a complex singlet extension of the Standard Model, with future LHC results.
We find that if another scalar is found, one can exclude a phase with a dark matter candidate in definite
regions of the parameter space, while predicting whether a third scalar to be found must be lighter or heavier. 
The first version of the code is publicly available and contains various generic core routines for tree level vacuum stability analysis, 
as well as implementations of collider bounds, dark matter constraints, electroweak precision constraints and tree level unitarity. 
}

\begin{document}
\maketitle




\section{Introduction}

The recent discovery of a scalar particle~\cite{ATLAS:2012ae, Chatrchyan:2012tx} at CERN's Large Hadron Collider (LHC) has boosted the activity in probing extensions of the scalar sector
of the Standard Model (SM). So far, the experimental results indicate that this scalar is compatible with the SM Higgs boson. However, many
of its extensions are also compatible with the present experimental results. In fact, we know that several of them
will never be completely disproved even if the Higgs has all the properties one expects from a SM Higgs. The limit where this occurs is known as
the decoupling limit and is characterised by a light Higgs with SM couplings to all other known particles while the remaining scalars are very heavy. It is possible
though that other scalars are waiting to be found at the LHC or that a more precise measurement of production cross sections and branching
ratios of the newly found 125 GeV scalar reveals meaningful deviations from the SM predictions. As one needs to be prepared to address that scenario
we have developed a new code to deal with the vacuum structure of scalar extensions of the SM. The code presented in this work,  \textsc{ScannerS}~\cite{ScannerS}, 
is intended to contribute to the analysis of the different scenarios that can be suggested by the LHC experiments.  

In the SM, electroweak symmetry breaking (EWSB) is achieved by one complex $SU(2)_L$ scalar doublet. Although the pattern of EWSB can be correctly
reproduced by this one scalar doublet, the SM is not able to accommodate a number of experimental results such as the existence of dark matter
or the measured baryon asymmetry of the universe. Adding a scalar singlet to the potential could provide a viable 
dark matter candidate~\cite{Silveira:1985rk, McDonald:1993ex, Burgess:2000yq, Bento:2000ah, Davoudiasl:2004be, Kusenko:2006rh, vanderBij:2006ne, He:2008qm,
Gonderinger:2009jp, Mambrini:2011ik, He:2011gc} 
as well as means
of achieving electroweak baryogenesis  by allowing a strong first-order phase transition during the era of EWSB~\cite{Menon:2004wv, Huber:2006wf, Profumo:2007wc,
Barger:2011vm, Espinosa:2011ax}. Although minimal, this extension provides a rich collider phenomenology leading to some distinctive signatures that
can be tested at the LHC~\cite{Datta:1997fx, Schabinger:2005ei, BahatTreidel:2006kx, Barger:2006sk, Barger:2007im, Barger:2008jx, O'Connell:2006wi,Gupta:2011gd}.

In this work we focus on the conditions on the parameters that lead to the existence of a global minimum in the scalar potential. However, contrary to previous works, we impose these conditions not to single out one particular  model (or phase), but rather use that information to distinguish between possible coexisting phases.
This way, we expect to identify properties of the models by classifying the possible phases as function of the parameter space point. We have also included in the code the most relevant theoretical and experimental constraints both from dark matter  and collider searches. Our goal is therefore to  
investigate the possibility of using measured experimental quantities (e.g. the mass and branching ratios of a new light scalar at the LHC), to automatically 
exclude one of the possible phases (e.g. a phase with a dark matter candidate) using the phase diagram of the model. To that purpose we scan over the entire parameter space
subject to the most relevant constraints and plot the results in projections that include physical quantities whenever possible.

As the ultimate goal of \textsc{ScannerS} is to be a tool that can be used for general scalar sectors, the core routines of the code can already be used for larger extensions of the scalar sector in terms of its field content. The core includes generic local minimum generation routines (with goldstone/flat direction identification as well as a-priori curved directions/symmetries) and tree level unitarity check routines. However, since the first version of the code has been tested extensively with the complex singlet models we present here, routines for testing electroweak observables can be used only for extensions with $n$-singlets, and global minimum and boundedness from below routines are defined on a model by model basis by the user. The same naturally applies for some of the analysis of experimental bounds (since such analysis will depend on the model that is being analysed) though some data tables from several experiments are generic and included in the core. We expect in the near future to automatise {\em global minimum} and {\em boundedness from below} routines to be in the core of the program. Details on the structure of the program and on how to use it can be found in~\cite{ScannerS}.  

The structure of the paper is the following. In Sect.~\ref{sec:models} we describe the scalar potential of the models we will study. 
We derive the conditions to be fulfilled for the minimum to be global at tree level for those specific models, based on symmetries, and classify the 
various phases for each model. In Sect.~\eqref{sec:scan_method} we address the problem of efficiently performing a scan of 
the parameters space in a more general perspective. We propose a method to generate a local minimum by generating vacuum expectation values (VEVs), 
mixings and masses uniformly, and obtaining some of the (dependent) couplings from the linear systems of equations which 
characterise the minimum. In Sect.~\ref{sec:TreeUni} we also discuss the generic implementation of tree level unitarity bounds.
In Sect.~\ref{sec:constraints_pheno} we implement constraints from various experimental sources, 
such as electroweak precision observables, LEP bounds, dark matter bounds, and constraints from Higgs 
searches at the LHC. Finally, we conclude with a discussion of our main results in Sect.~\ref{sec:conclusion}.

\section{The models}\label{sec:models}
We consider an extension of the scalar sector of the SM that consists on the addition of a complex singlet field
to the SM field content. 
Our starting point is the most general renormalisable model, invariant under a global $U(1)$ symmetry. We then consider two models with explicit breaking of this symmetry, but where $\mathbb{Z}_2$-type symmetries are preserved for one or two of the components of the complex scalar singlet. Each model is classified according to phases associated with the spontaneous symmetry breaking pattern of the vacuum.  

Considering the allowed parameter space, each model can exhibit more than one global minimum
with the correct pattern of EWSB for fixed couplings.
In these models,  the mixing matrix is usually lower dimensional, allowing for one or two unstable scalar bosons that
could be detected at the LHC and at future colliders, together with two or one dark matter candidates, respectively. However, if all $\mathbb{Z}_2$ symmetries are broken 
we can end up with three unstable scalars (and no dark matter candidate) complicating the signatures 
and making a clear identification of the Higgs boson a much more difficult task.
Our aim is to investigate the parameter space of these extensions as 
to identify the properties of the models which are not yet excluded by theoretical and experimental constraints. Furthermore, we want to investigate the 
possibility of, given a set of experimental measurements related to the scalar sector at the LHC, that we can automatically use the phase diagram of the model to 
exclude one of the phases, e.g. a phase with dark matter or a phase where all the scalars are mixed. 

The scalar field content of the model is as follows. There is the SM Higgs doublet $H$ which is a singlet under $SU(3)_C$
and is in the fundamental representations of $SU(2)_L\times U(1)_Y$, i.e. $\left(\mathbf{1},\mathbf{2},1/2\right)$.
We add a complex scalar field $\mathbb{S}=S+iA$ which is a singlet under the SM gauge group.
This is equivalent to adding two real singlet fields. Several models have been studied in the literature by
imposing special symmetries on this scalar sector. Our starting point are the models discussed in~\cite{Barger:2008jx}, 
where the Higgs potential has a global $U(1)$ symmetry in $\mathbb{S}$. Besides spontaneous breaking of this
symmetry by the vacuum, we can break it explicitly at various levels, by soft linear and quadratic terms which are
technically natural and do not generate other soft-breaking terms through renormalisation~\cite{Barger:2008jx}. 
The scalar potential is then (soft breaking terms in parenthesis)
\begin{equation}
V=\dfrac{m^2}{2}H^\dagger H+\dfrac{\lambda}{4}(H^\dagger H)^2+\dfrac{\delta_2}{2}H^\dagger H |\mathbb{S}|^2+\dfrac{b_2}{2}|\mathbb{S}|^2+
\dfrac{d_2}{4}|\mathbb{S}|^4+\left(\dfrac{b_1}{4}\mathbb{S}^2+a_1\mathbb{S}+c.c.\right) \, \, . \label{eq:V_general}
\end{equation}
We must ensure that the potential is bounded from below and in this simple model an analytic condition can be found easily.
Noting that the quartic operators of $V$ are quadratic forms in the real positive quantities $H^\dagger H$ and 
$|\mathbb{S}|^2$, positive definiteness of such forms is equivalent to boundedness from below and implies
\begin{equation}
\lambda>0 \; \; \; \wedge \; \; \; d_2>0 \; \; \; \wedge \; \; \; (\delta_2^2<\lambda d_2 \; \; \mathrm{if} \; \; \delta_2<0) \; .
\end{equation}
Before specifying the models, we note that the stationarity conditions are:
\begin{eqnarray}
\dfrac{\partial V}{\partial H^\dagger}&=&0\Leftrightarrow H\left(\dfrac{m^2}{2}+\dfrac{\lambda}{2}H^\dagger H+\dfrac{\delta_2}{2} |\mathbb{S}|^2\right)=0 \nonumber \\
\dfrac{\partial V}{\partial \mathbb{S}^\dagger}&=&0\Leftrightarrow \mathbb{S}\left(\dfrac{\delta_2}{2}H^\dagger H+\dfrac{b_2}{2}+\dfrac{d_2}{2}|\mathbb{S}|^2\right)+\dfrac{b_1^*}{2}\mathbb{S}^\dagger+a_1^*=0 \label{eq:MinCond} \; ,
\end{eqnarray}
which have no closed form solution in general. We will expand the Higgs doublet and the complex scalar field around the vacuum according to
\begin{equation}
\left<H\right>=\dfrac{1}{\sqrt{2}}\left(\begin{array}{c} 0 \\ v\end{array}\right) \; \;, \; \; \left<\mathbb{S}\right>=\dfrac{1}{\sqrt{2}}(v_S+i\,v_A)
\end{equation}
where the Higgs Vacuum Expectation Value (VEV) is $v=246\,{\rm GeV}$.

As previously stated, simpler versions of this model were already discussed in the literature. In most cases a specific model is 
singled out by imposing additional symmetries on the model. 
Because one of the main motivations for adding a scalar singlet to the SM is to provide a dark matter candidate,
models with one or two dark matter candidates are then analysed and confronted with experimental results.
Here, we are interested in applying our new tool~\textsc{ScannerS} to various versions of such models, while applying the latest experimental bounds, together with theoretical constraints. 
The points that pass all constraints during the scan are then classified according to the phase they are in (see classification below)
and plotted, whenever possible, in a physical projection of the parameter space, such as a mass of a new particle or a measurable rotation angle between group and mass eigenstates. A physical measurement allows us in  some cases to discriminate a phase with dark matter candidates from one with no dark matter candidates. Consequently, the phase which is realised for a given model, can in such cases be decided by experiment.

Theoretically, the independent models and their phases are classified according to their symmetry group and spontaneous breaking, respectively, as follows:
\begin{itemize}
\item
  {\em Model 0,} $U(1)$ symmetry with up to two dark matter candidates. This is obtained by imposing a $U(1)$ symmetry
   on the complex singlet field, which eliminates the soft breaking terms, thus $a_1=b_1=0$. There are two possible phases:
\begin{enumerate} 
\item $\left<\mathbb{S}\right>=0$ at the global minimum (symmetric phase). Then we have two degenerate dark matter particles. 
         In this scenario the model is equivalent to two independent real singlets of the same mass and quantum numbers.
\item 
   $\left<\mathbb{S}\right>\neq 0$ at the global minimum (broken phase). The $U(1)$ symmetry is spontaneously broken, 
   then there is an extra scalar state mixing with the Higgs and a (massless) Goldstone boson associated 
   with the breaking of the symmetry. Since the phase of the complex singlet is unobservable, without loss of 
   generality we can take $\left<S\right>\neq 0$ and $\left<A\right>=0$ for such phase and $A$ is the dark Goldstone particle.
   This phase is however strongly disfavoured by observations of the Bullet 
   Cluster~\cite{Randall:2007ph, Bento:2000ah, McDonald:2007ka, Barger:2008jx}. These observations
   can be used to constrain the mass of the dark matter particle as a function of the value of $\delta_2$.
   Hence, unless $\delta_2$ is vanishingly small, a zero mass dark matter particle is ruled out.   
\end{enumerate}
As one of the phases is not allowed, we will discard this model from our discussion.
\item
{\em Model 1,} $\mathbb{Z}_2\times \mathbb{Z}_2'$ symmetry with up to two dark matter candidates. 
This model is obtained by imposing a separate $\mathbb{Z}_2$ symmetry for each of the real components 
of the complex singlet. The $\mathbb{Z}_2$ symmetries imply that the soft breaking couplings are $a_1=0$ 
and $b_1\in \mathbb{R}$ (6 real couplings in the scalar potential \& no other couplings are generated 
through renormalisation). Specialising the minimum conditions~\eqref{eq:MinCond} we obtain the following 
qualitatively different possibilities for minima with $v^2 \neq 0$:
\begin{enumerate}
\item  $\left<\mathbb{S}\right>=0$, no mixing and two dark matter candidates (symmetric phase). 
\item $\left<S\right>=0$ or $\left<A\right>=0$, one of the singlet components mixed with the
 Higgs doublet and one dark matter candidate (spontaneously broken phase). One can show, by noting that swapping $S \leftrightarrow A$ 
 only changes the sign of $b_1$, that without loss of generality we can take $\left<A\right>=0$, while 
 still covering the full parameter space (this is so because $b_1\in \mathbb{R}$, and the potential only depends on squares of the VEVs). 
 This is true in our scans only because we will adopt the strategy of first generating a locally viable minimum and, only after, to check all possibilities for minima below the one generated.
\end{enumerate}

\item
{\em Model 2}, One $\mathbb{Z}_2'$ symmetry with up to one dark matter candidates. This is obtained by imposing a $\mathbb{Z}_2'$ 
symmetry on the imaginary component $A$. Then the soft breaking couplings must be both real, i.e. $a_1\in\mathbb{R}$
 and $b_1\in\mathbb{R}$. Looking at the minimum conditions we find the following cases:
\begin{enumerate}
\item $\left<A\right>=0$, i.e. mixing between $h$ (SM Higgs doublet fluctuation) and $S$ only (symmetric phase). In this case we can take $S\in \mathbb{R}^+$ as long as $a_1$ runs through positive and negative values.

\item $\mathbb{S}\neq 0$, i.e. both VEVs non-zero and mixing among all fields (broken phase).
\end{enumerate}

\end{itemize}
To summarise this classification, we show in table~\ref{tab:phases} a list of the three possible models (labelled by their symmetry group), 
with a description of the particle content of the two possible phases (symmetric or broken) as well as the VEV/symmetry breaking pattern. 
\begin{table}
\begin{center}
\begin{tabular}{||  c |}
	
\hline			
  Model  \\
\hline 
\hline	
$\mathbb{U}(1)$  \\ $\phantom{\mathbb{U}(1)}$\\
\hline
\hline	
$\mathbb{Z}_2\times \mathbb{Z}_2'$  \\ $\phantom{\mathbb{U}(1)}$\\
\hline
\hline	
$\mathbb{Z}_2'$ \\ $\phantom{\mathbb{Z}_2\times \mathbb{Z}_2'}$ \\
\hline  
\end{tabular}
\begin{tabular}{| c |}
	
\hline			
  Phase  \\
\hline	
\hline		
  Higgs+2 degenerate dark \\
\hline			
  2  mixed + 1 Goldstone \\
\hline
\hline			
   Higgs + 2  dark  \\
\hline			
  2 mixed + 1 dark \\
\hline		
\hline		
  2 mixed + 1  dark \\
\hline		
  3 mixed \\
\hline
\end{tabular}
\begin{tabular}{| c |}
	
\hline			
  VEVs at global minimum \\
\hline	
\hline		
   $\left<\mathbb{S}\right>=0$ \\
\hline			
    $\left<A\right>=0$ ($\cancel{\mathbb{U}}(1)\rightarrow \mathbb{Z}_2'$) \\
\hline
\hline	
   $\left<\mathbb{S}\right>=0$ \\
\hline			
   $\left<A\right>=0$ ($\cancel{\mathbb{Z}}_2\times \mathbb{Z}_2'\rightarrow \mathbb{Z}_2'$ ) \\		
\hline		
\hline	
   $\left<A\right>=0$ \\
\hline			
  $\left<\mathbb{S}\right>\neq 0$ ($\cancel{\mathbb{Z}}_2'$ ) \\
\hline
\end{tabular}

\end{center}
\caption{Phase classification for the three possible models.}
\label{tab:phases}
\end{table}

Once we have picked one of the cases above, we have to check for all possible minima by evaluating the potential at all possible stationary points. A list of the possible cases is provided in appendix~\ref{stationary}.

We have chosen to start the studies with our new tool with these models, because they already have a great 
diversity of physically different phases (which provide some interesting results -- see Sect.~\ref{sec:conclusion}), 
while allowing us to test the routines of the code we propose to develop. We have reproduced several results presented
in the literature with a very good agreement. In the next sections, we describe the \textsc{ScannerS} 
scanning strategy and the main tree level theoretical constraints to generate a stable vacuum with the correct symmetry breaking pattern.

\section{The scanning method}\label{sec:scan_method}

To implement the scan of the parameter space of the models, we have developed a dedicated tool \textsc{ScannerS} 
which can be used for more generic potentials. Our method is based on a strategy of reducing as many steps 
as possible to linear algebra, since these are computationally less expensive. Before describing the method in detail,
 let us describe the general idea. 

A possible way of performing the scan for a generic scalar potential (strategy 1), and determining the 
spectrum of scalars is: i) scan the couplings $\lambda_a$ uniformly in chosen ranges, ii) determine 
all stationary points by solving the (non-linear) stationarity conditions, iii) check if any are minima
and choose the global one iv) if yes, accept the point, compute the mass matrix, diagonalise and check
 if the masses of the scalars ($m_i^2$) and mixing matrix ($M_{ij}$) are consistent with the symmetry 
 breaking imposed. This method contains two steps which are quite expensive, computationally, which 
 are executed before we know if the minimum has the desired properties. The first is the determination 
 of the stationary points, which in general is a problem of finding the solutions of a polynomial system 
 of equations in the VEVs. The second, not as expensive, is the diagonalisation of the
  mass matrix at the global minimum.

Our alternative strategy (strategy 2) relies on two observations. First,  a generic scalar potential 
$V(\phi_i)$ (where the $\phi_i$ are the real fields used to decompose the scalar 
multiplets/singlets into real components), is a linear form in the couplings $\lambda_a$
\begin{equation}
V(\phi_i)=V_a(\phi_i)\lambda_a \; , \label{eq:Vlinear}
\end{equation}
where (for renormalisable models) $V_a(\phi_i)$ are monomials in the fields, of degree 
up to four. Second, in strategy 1, the couplings $\lambda_a$  are taken as independent parameters, and the VEVs, masses and mixing parameters are determined. However, any independent set of parameters is
 equally good to label a certain point in parameter space, so if instead we  use VEVs, masses and mixing matrix elements as 
 the parameters to be scanned over, then the determination of dependent couplings $\lambda_a$ becomes a linear 
 algebra problem due to the linearity property of~\eqref{eq:Vlinear}. Furthermore, this set of parameters (VEVs, masses and mixings) is
  more directly related to physical properties of the scalar states, so it is a more natural choice of parametrisation. 

\subsection{Generation of a local minimum}
The details of the method we use are as follows. We first want to express the Lagrangian in terms of physical propagating scalar degrees of freedom $H_i$ with masses $m_i$, which are the fluctuations around a minimum with VEVs $v_i$. The most general expansion of the original fields such that the kinetic terms remain canonical is then
\begin{equation}\label{eq:general_expansion}
\phi_i=v_i+M_{ij}H_j
\end{equation} 
where $M_{ij}$ is a generic rotation matrix, and we have assumed that the original kinetic terms for the $\phi_i$ fields are canonically normalised. The (linear) stationarity conditions (vacuum conditions) are
\begin{eqnarray}
\left.\dfrac{\partial V}{\partial \phi_i}\right|_{v_i} = 0\Leftrightarrow \left<\partial_i V\right>_a\lambda_a&=&0 \label{eq:gen_linear}
\end{eqnarray}
where we have defined the derivative of the $a$-th operator with respect to the field $\phi_i$ evaluated at the VEVs by $\left<\partial_i V\right>_a$. Such object is as matrix with indices $\left\{i,a\right\}$ acting on a vector of couplings $\lambda_a$.  These conditions are independent of the mixing matrix. If we choose values for the VEVs (by scanning over an interval or fixing some of them such as the SM Higgs VEV), this system of equations can be analysed as to identify a sub-set of couplings to be eliminated in favour of the remaining couplings (let them be grouped in a sub-vector $\lambda_{a_1}$ where the sub-index $a_1$ runs only over dependent couplings). Otherwise, if the system is over-determined we must reject such choice of VEVs.

The dependent couplings $\lambda_{a_1}$ can be eliminated in favour of the remaining ones, whose subset we denote $\lambda_{a_2}$\footnote{Note: One can always favour keeping couplings of lower dimensional operators as independent, since they are usually easier to interpret in terms of observable quantities. Or alternatively they can be ordered according to any other criterion.}. This elimination is represented in a matrix form as follows
\begin{equation}
\lambda_{a_1}=\Lambda_{a_1a_2}\lambda_{a_2} 
\end{equation}
where the matrix $\Lambda_{a_1a_2}$ can be found numerically through Gaussian elimination using the system~\eqref{eq:gen_linear}. This procedure amounts to replacing the $\lambda_{a_1}$ couplings by the VEVs that have been scanned over. 

The vacuum conditions can be implemented in any VEV of some operator $\mathcal{O}$ acting on the potential, using the matrix $\Lambda_{a_1a_2}$:
\begin{equation}
\left<\mathcal{O}V\right> = \left(\left<\mathcal{O}V_{a_1}\right>\Lambda_{a_1a_2}+\left<\mathcal{O}V_{a_2}\right>\right)\lambda_{a_2} \equiv \widehat{\mathcal{O}}V_{a_2}\lambda_{a_2} \; , \label{eq:def_vevreduced}
\end{equation}
where in the last step we have defined the ``VEV reduced'' action of $\mathcal O$ on the potential, $\widehat{\mathcal{O}}V_{a_2}$, which contracts (linearly) with the sub-vector of couplings $\lambda_{a_2}$. 

The second step is to impose that the stationary point is a minimum, which is done by assuming non-negative masses squared. Thus we write the quadratic derivative conditions which involve the masses and mixing matrix elements, i.e.
\begin{eqnarray}
\left.\dfrac{\partial^2 V}{\partial H_i \partial H_j}\right|_{H_i=0}=M_{ik}M_{jl}\left<\widehat{\partial}^2_{kl}V\right>_{a_2}\lambda_{a_2}&=&\delta_{ij}m_{\hat{i}}^2 \nonumber\\
\Leftrightarrow \mathbf{M}^T\left<\widehat{\mbox{\boldmath$\partial$} }^2V\right>_{a_2}\mathbf{M} \,\lambda_{a_2}&=&\mathbf{Diag}(m_i^2) \label{eq:Matrix}
\end{eqnarray}
where we have used Eqs.~\eqref{eq:general_expansion} and \eqref{eq:def_vevreduced}, and in the last line we have replaced latin indices by a bold face matrix notation. The hatted index $\hat i$ denotes no summation over $i$. For a given model, the mass matrix before diagonalisation 
\begin{equation}
\left<\widehat{\mbox{\boldmath$\partial$} }^2V\right>_{a_2}\lambda_{a_2} \; ,
\end{equation} may have eigen-directions which are independent of the values of the couplings $\lambda_{a_2}$. In such case we say that there are ``a priori'' eigen-directions at the minimum. This can be tested by finding the eigen-vectors of a matrix $\left<\widehat{\mbox{\boldmath$\partial$} }^2V\right>_{a_2}$ for fixed $a_2$, and then check which eigen-vectors remaining eigen-vectors of the other matrices with different $a_2$. We may then find ``a priori'' flat eigen-directions (which would be Goldstone bosons) or curved eigen-directions (which would be massive particles)\footnote{This procedure can be continued until all ``a priori'' eigen-directions are found.}. Once those directions are identified, the form of the mixing matrix is restricted to a block diagonal form. The flat directions fix the mass of the corresponding particles to zero regardless of $\lambda_{a_2}$, so effectively they eliminate a set of conditions from system~\eqref{eq:Matrix}. For each curved eigen-directions, there is a condition with the mass squared of the particle on the right hand side. For the block where mixing occurs, since the mixing matrix is symmetric, we have $n(n+1)/2$ conditions ($n$ is the dimension of the block). Along the diagonal of the block, $n$ conditions contain a mass squared on the right hand side. The $n(n-1)/2$ conditions off the diagonal conditions contain a zero on the right hand side. 

At this stage the undetermined parameters are: the $\lambda_{a_2}$ couplings; the mixing matrix elements $M_{ij}$; and the physical masses $m_i^2$. Because we want to avoid solving non-linear equations, we choose to generate first the mixing matrix uniformly\footnote{A rotation matrix can be generated with uniform probability with respect to the Haar measure, by generating its entries with a Gaussian distribution and then performing a QR decomposition to extract it~\cite{press_numerical_1992}.}. Then we are left with a set of conditions relating $\lambda_{a_2}$ with the masses $m_i^2$, which can be re-arranged as a homogeneous system. If we define the vector of parameters which are still undetermined by $\mathbf{v^T}=(\lambda_{a_2},m_i^2)$ such homogeneous system is
\begin{equation}
\mathbf{D}\mathbf{v}=0
\end{equation}
where the matrix $\mathbf{D}$ is read out from~\eqref{eq:Matrix}. Using again Gaussian elimination we can solve for a sub-set of parameters of $\mathbf{v}$, as a function of another subset which we are left to scan over (if the system is over-determined we must reject the vacuum).

With this procedure, we end up generating a point in parameter space, and all the properties of the physical states are determined (such as masses and mixing matrices), having avoided the problem of solving a system of non-linear equations. The price to pay is that we have delayed checking if the minimum is global. However, the advantage of this procedure is that we do not spend any time in points of parameter space where there is no stationary point with the correct properties. Furthermore, we can add to this procedure more constraints on the local minimum (if they are computationally quick to check), before checking if it is global. If we use strategy 1, the first steps involve a computationally intensive non-linear problem, which will be wasted each time a point in parameter space is rejected, whereas with strategy 2 we generate a local minimum with the desired properties quickly and only then do we have to perform the computationally intensive steps.

\subsection{Tree level unitarity}
\label{sec:TreeUni}
An important constraint on the region of parameter space to be scanned over is given by imposing tree level unitarity at high energies. This was pioneered by Lee, Quigg and Thacker in~\cite{Lee:1977eg}. It was shown that the Goldstone high energy theorem applies and all we need is to compute all the scalar quartic interaction amplitudes, describing $2\rightarrow 2$ processes between any (suitably normalised) scalar two particle states. The basis of such states for $N$ real scalars is
\begin{equation}
\left(\begin{array}{c}
\ldots\\
\ldots\\
\ldots \\
\left|\Phi_i\right> \vspace{1mm}\\
\ldots\\
\ldots
\end{array}\right)\equiv\left(\begin{array}{c}\tfrac{1}{\sqrt{2!}}\left|\phi_1\phi_1\right>\\
\ldots \\
\tfrac{1}{\sqrt{2!}}\left|\phi_N\phi_N\right> \vspace{1mm}\\
\left|\phi_1\phi_2\right> \\
\ldots\\
\left|\phi_{N-1}\phi_N\right>
\end{array}\right)
\end{equation}
where we have emphasised the $\frac{1}{\sqrt{2!}}$ term which arises from the normalisation of harmonic oscillator (indistinguishable) 2-particle states, and the second block contains all the pairings of different field states. Then, one constructs an s-wave amplitude matrix 
\begin{equation}
 a^{(0)}_{ij}=\dfrac{1}{16\pi}\left<\Phi_i\right|i\mathbf{T}^{(0)}\left|\Phi_j\right>=\dfrac{1}{16\pi}\sum_{a_4}\dfrac{P_{a_4}}{\sqrt{n_i!n_j!}} \lambda_{a_4}\label{eq:a0mat}
\end{equation} 
where $\mathbf{T}^{(0)}$ is the $J=0$ contribution to the usual transition matrix. In the second step we have used the Feynman rules for the scalar sector to express each matrix element as a sum over quartic vertices labelled by the index $a_4$. If the $\left|\Phi_i\right>$ state contains identical particles, $n_i=2$  otherwise it is one. $P_{a_4}$ is defined as the product of the factorials of the powers of each field in the corresponding monomial $V_{a_4}(\phi)$. For a generic scalar potential,~\eqref{eq:a0mat} yields real coefficients. Tree level \mbox{s-wave} unitarity, is then implemented by requiring that the absolute value of each eigenvalue of $a^{(0)}_{ij}$ is smaller than $1/2$. This matrix can be computed efficiently at each point of parameter space, and diagonalised, as to check if unitarity is preserved. For the complex singlet model we are considering, one can write down exactly the most restrictive conditions which are
\begin{equation}
|\lambda|\leq 16\pi \;\wedge \; |d_2|\leq 16\pi \; \wedge \; |\delta_2|\leq 16 \pi \; \wedge \; \left|\frac{3}{2}\lambda+d_2\pm \sqrt{\left(\frac{3}{2}\lambda+d_2\right)^2+d_2^2}\right|\leq 16\pi \; ,
\end{equation}
which correctly reduces to the SM tree level unitarity bound when $d_2=\delta_2=0$.

\section{Constraints and phenomenological potential}\label{sec:constraints_pheno}

\subsection{Electroweak precision observables}

Another source of constraints comes from electroweak precision observables. Here we focus on the $S,T,U$ variables~\cite{Maksymyk:1993zm,Peskin:1991sw}. For an extended sector with scalar singlets coupling only to the Higgs doublet, the only extra contributions compared with the SM are in the self energies $\Pi_{ZZ}(q^2)$ and $\Pi_{WW}(q^2)$ for the $Z$ and $W$ particles respectively. Using the general expressions in~\cite{Barger:2007im}, one can show that the variations with respect to the SM contributions from a single Higgs doublet are
\begin{eqnarray}
\Delta S&=&\Delta\left[\dfrac{1}{\pi}\sum_j\left(M_{hj}\right)^2\left\{f\left(\tfrac{m_j^2}{M_Z^2}\right)-g\left(\tfrac{m_j^2}{M_Z^2}\right)\right\}\right]\\
\Delta T&=&\Delta\left[\frac{1}{4\pi s_W^2}\sum_j\left(M_{hj}\right)^2\left\{g\left(\tfrac{m_j^2}{M_W^2}\right)-\frac{1}{c_W^2}g\left(\tfrac{m_j^2}{M_Z^2}\right)\right\}\right]\\
\Delta U&=&\Delta\left[\frac{1}{\pi}\sum_j\left(M_{hj}\right)^2\left\{f\left(\tfrac{m_j^2}{M_W^2}\right)-g\left(\tfrac{m_j^2}{M_W^2}\right)-f\left(\tfrac{m_j^2}{M_Z^2}\right)+g\left(\tfrac{m_i^2}{M_Z^2}\right)\right\}\right]\
\end{eqnarray}
where the terms inside the brackets are to be evaluated with the values of the mixing matrix elements $M_{hj}$ that 
represent the mixing between the SM Higgs and the new singlet field components. The masses $m_i^2$
correspond to the new scalar physical eigenstates. 
The functions $f(Y)$ and $g(Y)$ are defined as 
\begin{eqnarray}
f(Y)&=&-\dfrac{1}{4}Y\log Y+\int_0^1dx \left[-\dfrac{3}{2}+\left(\dfrac{Y}{2}+1\right)x-\frac{x^2}{2}\right]\log\left((1-x)^2+xY\right) \\
g(Y)&=&\dfrac{Y}{4}\left(\dfrac{\log Y}{1-Y}-\dfrac{1}{2}\right) \; .
\end{eqnarray}
Denoting the variation of the three observables in a vector $\Delta \mathcal{O}_i\equiv \mathcal{O}_i-\mathcal{O}^{SM}_i\rightarrow \left(\Delta S,\Delta T,\Delta U\right)$, then consistency with the electroweak fit within a 95\% C.L. ellipsoid is implemented by requiring 
\begin{equation}
\Delta \chi^2\equiv\sum_{ij}\left(\Delta \mathcal{O}_i-\Delta \mathcal{O}^{(0)}_i\right){\left[(\sigma^2)^{-1}\right]}_{ij}\left(\Delta \mathcal{O}_j-\Delta \mathcal{O}^{(0)}_j\right)<7.815
\end{equation}
where the covariance matrix is defined in terms of the correlation matrix, $\rho_{ij}$, and the standard deviation, $\sigma_i$, of each parameter through $\left[\sigma^2\right]_{ij}\equiv \sigma_i\rho_{ij}\sigma_j$. To test these observables, we have used results for the SM global fit of the Gfitter collaboration~\cite{Baak:2011ze} with a Higgs mass\footnote{Note: The latest fit central values differ only very slightly to these numbers so we do not expect a noticeable difference. } $m_h=120\, {\rm GeV}$ and a top mass $m_t=173.1\, {\rm GeV}$. Their results are:
\begin{eqnarray}
S  &=& 0.02 \pm 0.11 \Rightarrow \Delta S^{(0)}= -0.03 \pm 0.11\nonumber\\ 	
T  &=& 0.05 \pm 0.12 \Rightarrow \Delta T^{(0)}= 0.03 \pm 0.12  \\	
U &=& 0.07 \pm 0.12 \Rightarrow \Delta U^{(0)}= 0.06 \pm 0.12 \nonumber
\end{eqnarray}
where in the last line we have used the values  $S_{SM}=0.05$, $T_{SM}=0.02$, $U_{SM}=0.01$ computed for the SM contribution from a single Higgs with $m_h=120 \, {\rm GeV}$, using the expressions in~\cite{Barger:2007im} specialised to the SM. The correlation matrix is
\begin{equation}
\rho_{ij}=\left(\begin{array}{ccc} 1&0.879 &-0.469\\0.879&1 &-0.716 \\-0.469&-0.716 &1\end{array}\right) \; .
\end{equation}
\subsection{LEP and LHC bounds}

An important set of experimental constraints comes from collider searches for the Standard Model Higgs boson. 
The standard strategy is to compute, for each search channel, the predicted signal strength defined as
\begin{equation}
\mu_i =\dfrac{\sigma_{\rm New}(H_i){\rm Br}_{\rm New}\left(H_i\rightarrow X_{\rm SM}\right)}{\sigma_{\rm SM}(h_{SM}){\rm Br}_{\rm SM} \left(h_{SM}\rightarrow X_{\rm SM}\right)} \, \, 
\end{equation}
where $\sigma_{\rm New}(H_i)$ and $\sigma_{\rm SM}(h_{SM})$ are the production cross sections of $H_i$ and the SM Higgs respectively, both evaluated at the mass of $H_i$;
$ {\rm Br}_{\rm New}\left(H_i\rightarrow X_{\rm SM}\right)$ is the $H_i$ branching ratio (BR) to SM particles and 
${\rm Br}_{\rm SM} \left(h_{SM}\rightarrow X_{\rm SM}\right)$ is the SM Higgs BR to SM particles evaluated at the mass of $H_i$.
In the models we are considering, the scalars couple to the SM particles always through the same combination $h=M^T_{hi}H_i=M_{ih}H_i$. Therefore, 
 both the production cross sections and the decay widths are just rescaled by the factor $M_{ih}^2$. We can then write $\mu_i$ as
\begin{equation}\label{eq:mu_xSM}
\mu_i = M_{ih}^2 \dfrac{{\rm Br}_{\rm New}\left(H_i\rightarrow X_{\rm SM}\right)}{{\rm Br}_{\rm SM}\left(h_{SM}\rightarrow X_{\rm SM}\right)} \, .
\end{equation}
However, because there are new particles involved, the ratio of BRs is now
\begin{equation}
\dfrac{{\rm Br}_{\rm New}\left(H_i\rightarrow X_{\rm SM}\right)}{{\rm Br}_{\rm SM}\left(h_{SM}\rightarrow X_{\rm SM}\right)}=
\dfrac{M_{ih}^2\Gamma(h_{SM}\rightarrow  X_{\rm SM})}{M_{ih}^2 \Gamma(h_{SM}\rightarrow  X_{\rm SM})+\sum\Gamma(H_i\rightarrow new scalars)} \, ,
\end{equation}
where the term $\sum\Gamma(H_i\rightarrow new scalars)$ is only present when the channels for which the SM Higgs decays to the new scalars are open, and once again $\Gamma(h_{SM}\rightarrow  X_{\rm SM})$ denotes the SM Higgs width evaluated at the mass of $H_i$.
We only consider two-body final states for the Higgs boson decaying to other scalars. Then, when kinematically allowed, the decay widths for a process of the type $H_i\rightarrow H_j H_j$  and $H_i\rightarrow H_j H_k$  are given respectively by
\begin{eqnarray}
\Gamma\left(H_i\rightarrow H_j H_j\right)&=& \dfrac{g^2_{ijj}}{32\pi m_{i}}\sqrt{1-\dfrac{4m_{j}^2}{m_{i}^2}} \\
 \Gamma\left(H_i\rightarrow H_j H_k\right)&=& \dfrac{g^2_{ijk}}{16\pi m_{i}}\sqrt{1-\dfrac{(m_{j}+m_{k})^2}{m_{i}^2}}\sqrt{1-\dfrac{(m_{j}-m_{k})^2}{m_{i}^2}}\, ,
\end{eqnarray}
where $g_{ijj}$, $g_{ijk}$ are the coupling strengths between the corresponding scalars $i,j,k$ and $m_{j}$ is the mass of the scalar state $H_j$. 

The combined LEP data \cite{Barate:2003sz} from the four LEP collaborations sets a 95 \% confidence level upper bound on the $H_i ZZ$ 
coupling in non-standard models relative to the same coupling in the SM, through the quantity
\begin{equation}
\chi_i^2 = \left(\frac{g^{BSM}_{H_i ZZ}}{g^{SM}_{HZZ}}\right)^2 \times {\rm Br}_{\rm New} (H_i \rightarrow ZZ)=\mu_i \, {\rm Br}_{\rm SM}\left(h_{SM}\rightarrow ZZ\right)
\end{equation} 
where $g_{H_i ZZ}^{BSM}$ designates the non-standard $H_i ZZ$ coupling while $g_{HZZ}^{SM}$ stands for the corresponding SM coupling. In the last line we
 have related this quantity to the signal strength $\mu_i$ for this particular channel, as defined in Eq.~\eqref{eq:mu_xSM}. 
We apply the LEP limits for each scalar mass eigenstate $H_i$ in the $b {\bar b}$ and $\tau^+ \tau^⁻$ decay channels separately.

With the recent discovery of a Higgs like boson by the ATLAS and CMS collaborations,
we now have a very strong constraint on BSM models. In any extension of the SM
one of the scalars is bound to have a mass of approximately 125 GeV. At the same time
both LHC experiments have also constrained any new scalar couplings to the SM
particles, that is, they have provided us an exclusion region for $\mu_i$ as function of the
mass of the scalar $H_i$. These bounds are also included in the program and the details
are as follows. We assume a Higgs boson with mass $m_h=125 \, {\rm GeV}$, and allow for a signal strength 
$\mu_i$ in the interval $1.1\pm 0.4$~\cite{CSearch}.  
In the mass regions where a scalar particle is excluded,  we apply the 95 \% CL combined ATLAS upper limits on
$\mu_i$ as a function of $m_{i}$~\cite{CSearch} for all other non 125 GeV scalars, as long as their production is allowed at the LHC. 

\subsection{Dark matter experimental bounds}
We have applied the limits on the total dark matter relic density from WMAP (Wilkinson Microwave Anisotropy Probe) 7-year measurements of cosmic microwave background (CMB) anisotropies~\cite{PDG} 
\begin{equation}
\Omega_{cdm} h^2= 0.112 \pm 0.006
\end{equation} 
where $h$ is the Hubble constant ($h = 0.704 \pm 0.025$ in units of 100 km/Mpc/s).
We have calculated numerically the relic density for the DM candidate with the software package micrOMEGAS~\cite{Belanger:2006is,Belanger:2010gh}. 
We have allowed the contribution from the dark matter candidate in each model
to be below the limit for the relic density, taking into account that there could be another DM contributor. 

Another important constraint on these models is the elastic cross section from DM scattering off nuclear targets.
The most recent direct detection DM experiments have placed limits in the spin-independent (SI)
scattering cross section ($\sigma_{SI}$) of weakly interacting massive particles (WIMPS) on nucleons.
The most restrictive upper bound on the SI elastic scattering of a DM particle is the one from XENON100~\cite{Aprile:2012nq}.
In \textsc{ScannerS} we have included  the limits of $\sigma_{SI}$ as a function of the dark matter mass as presented in~\cite{Aprile:2012nq}.

In these singlet models, the scattering cross section of the dark matter candidate with a proton target is given by~\cite{Barger:2010yn} 
\begin{equation}
\sigma_{SI} = \frac{m_p^4}{2\pi v^2}\sum_i\dfrac{1}{(m_p + m_{i})^2} \left[\sum_j \frac{ M^2_{j h} g_{i j j}}{m^2_{j}}\right]^2 \left( f_{pu} + f_{pd} + f_{ps} + \frac{2}{27} (3 f_G)\right)^2 
\end{equation}
where $m_p$ is the proton mass, $v=246$ GeV is the SM-Higgs VEV and $f_{pi}$ are the proton matrix elements with central values~\cite{Ellis:2000ds}:
\begin{equation}
f_{pu} =0.020 \quad f_{pd} =0.026 \quad f_{ps} =0.118 \quad f_G = 0.836 \, . 
\end{equation}
The dominant contribution for SI scattering comes from t-channel scalar exchange. We note that the cross-sections for scattering off protons is very similar
to the one for neutrons. Therefore we present only the results for the scattering off proton targets. Finally we should mention that we have used micrOMEGAS~\cite{Belanger:2006is} 
to perform an independent calculation of the cross sections and that we have found a very good agreement with our result. 
We have written the cross section expressions for the models considered in Appendix~\ref{SIcs}.


\section{Discussion}
\label{sec:conclusion}
\begin{figure}
\begin{center}
\includegraphics[clip=true,scale=0.5,trim = 3.8cm 5.2cm 4cm 9.8cm]{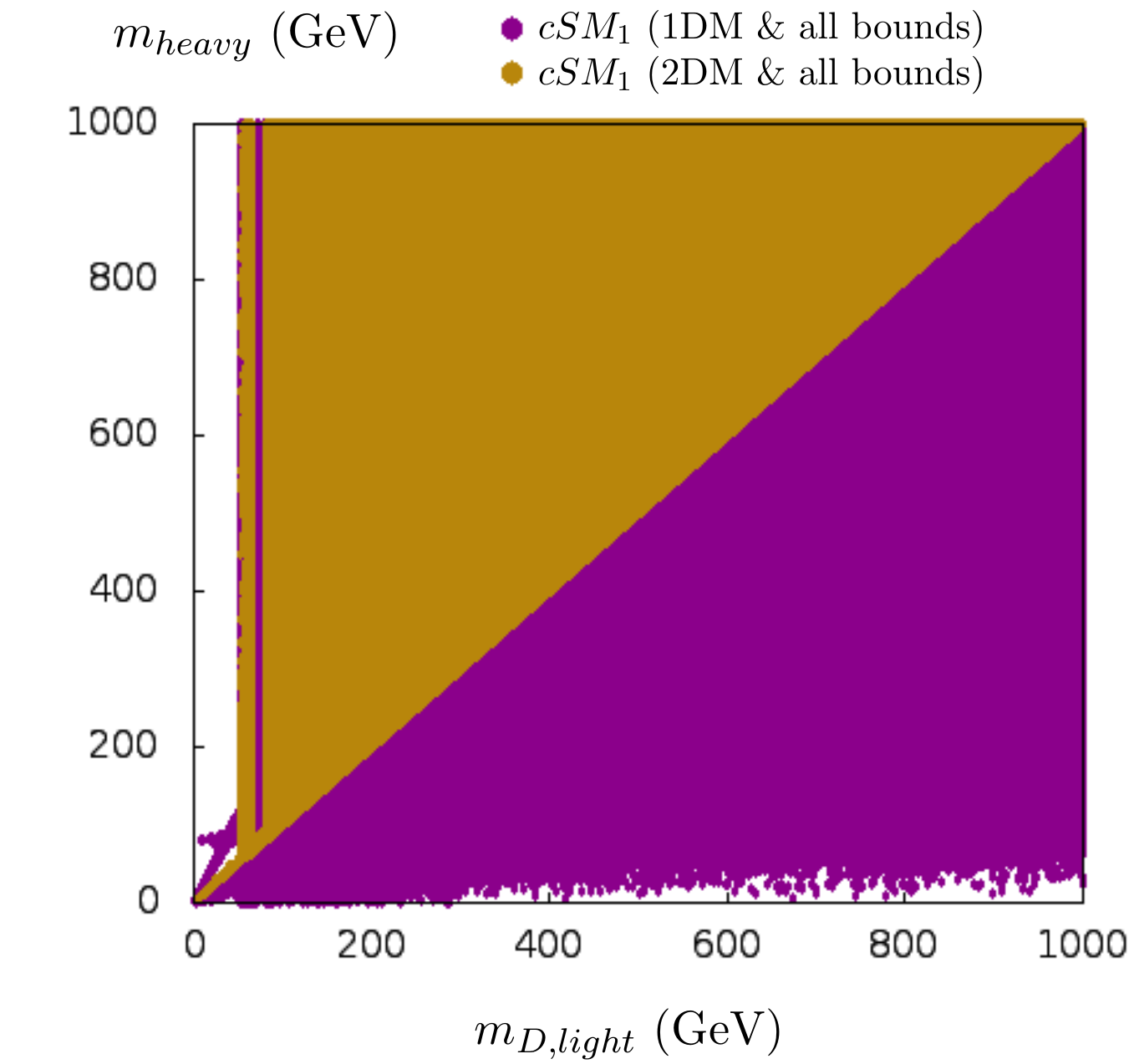}
\includegraphics[clip=true,scale=0.5,trim = 2.5cm 5.2cm 4cm 9.8cm]{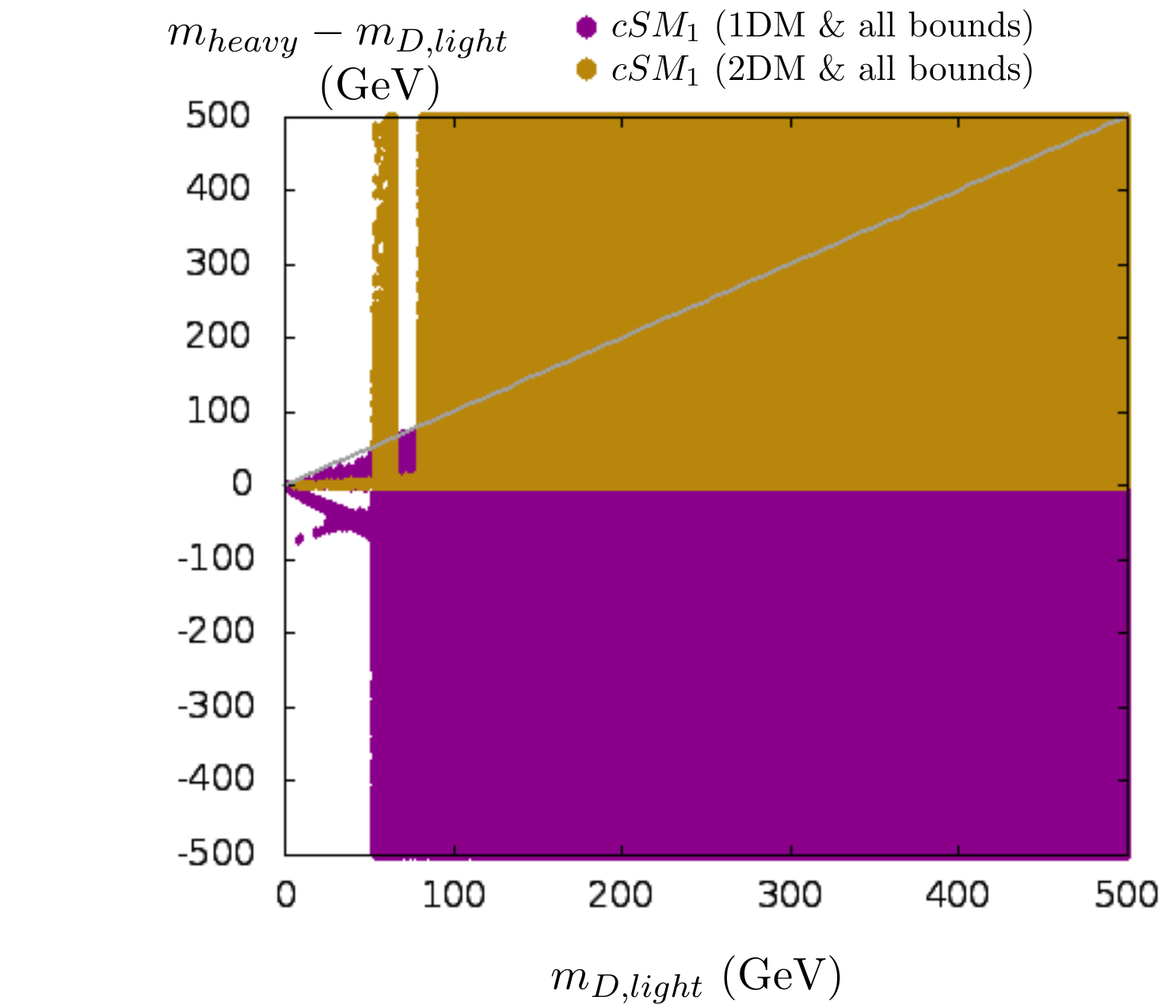}
\end{center}
\caption{\label{fig_phase_cSM1}{\em Phase diagram for model 1 ($cSM_1$)}: Scatter plots of two projections 
of the parameter space points obtained for a wide scan of the two phases of model 1. In each plot, 
we have overlaid the points of the phase which does not cover most of the plane, on top of the other phase which does.
{\em Left}: $m_{heavy}$ as a function of $m_{D,light}$ where $m_{D,light}$ is the mass of the lightest dark matter particle 
and $m_{heavy}$ is the mass of the the other scalar (dark or not); {\em Right:} $m_{heavy}-m_{D,light}$ as a function of $m_{D,light}$. }
\end{figure}

In this section we analyse the results of full scans over all parameter space, for the various phases of each model. We focus on models 1 and 2, which we denote by $cSM1$ and $cSM2$ respectively. We have performed two main scans for each model. One with a smaller hyper-cubic box in parameter space to allow for a better resolution of the region being scanned over, and another wider scan to check which boundaries did not change significantly. Unless stated otherwise, we always use the wider scan\footnote{The exact ranges of the scans that we have performed are presented in appendix~\ref{range}.}. We will also present a scan with some of the constraints removed to clarify
the appearance of certain boundaries  in the allowed regions of parameter space.

We start by presenting some results for model 1 ($cSM1$)  where we label the phase with two dark matter candidates as ``2DM'' and the phase with only one dark matter candidate by ``1DM''. The key of all figures contains an extra label for each colour of the points, indicating whether all the bounds/constraints discussed in previous sections have been included, and when not, the corresponding constraints that have been removed, are indicated.

In Fig.~\ref{fig_phase_cSM1} we show two particular projections in parameter space which show regions which are exclusive of a given physical phase. On the left panel we display $m_{heavy}$ as a function of $m_{D,light}$ where $m_{D,light}$ is the mass of the lightest dark matter particle and $m_{heavy}$ is the mass of the the other non-SM scalar;  on the right we show $m_{heavy}-m_{D,light}$ as a function of $m_{D,light}$. Note that the heavy scalar
is in fact a dark matter candidate in the phase ``2DM''.

There are regions exclusive to the phase with two dark matter candidates, and also very small regions where just one dark matter state is possible. 
A measurement of the mass of a dark matter candidate could  give us a hint on the phase the model is in. However, even if it is possible
in some cases to distinguish between phases with actual physical measurements, it is rather hard to discriminate between phases in model 1,
even more so because in some regions it requires independent measurements of the properties of a dark matter candidate and an unstable scalar mixing with the Higgs.

\begin{figure}
\begin{center}
\includegraphics[clip=true,scale=0.507,trim = 3.6cm 5.3cm 4cm 8cm]{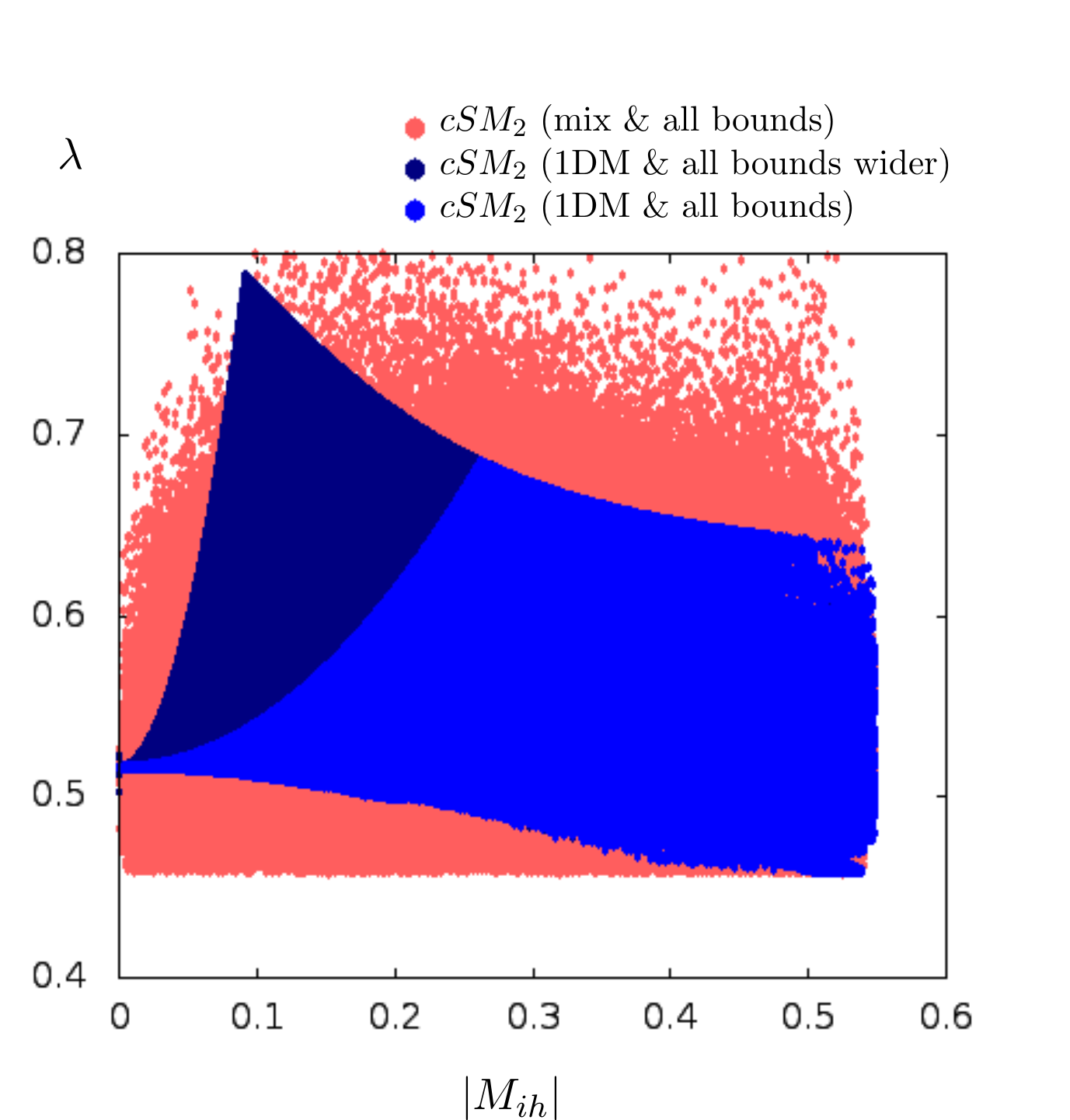}\includegraphics[clip=true,scale=0.507,trim = 2.7cm 5.3cm 4cm 8cm]{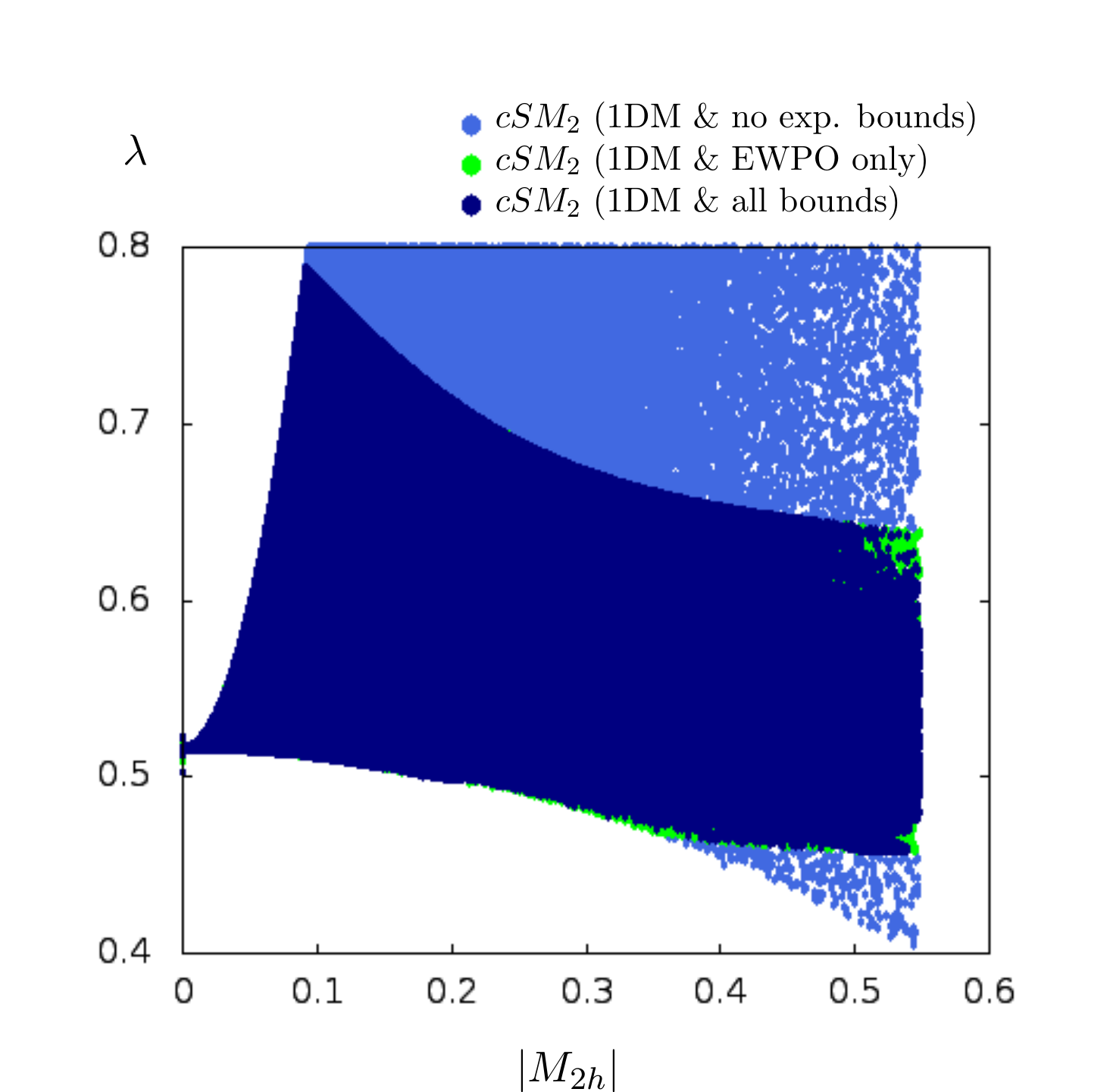}
\end{center}
\caption{\label{fig_cSM2F1}{\em Phase diagram for model 2 ($cSM_2$)}: 
Scatter plot of various projections of the parameter space points of the two phases of model 2. On the left we plot $\lambda$ (the doublet quartic coupling)
as a function of $|M_{ih}|$ which is the mixing matrix element with the SM Higgs doublet component of (any of) the new mixed scalar(s) which is not the $125$~GeV scalar.
In this case, we show a standard and a wide scan for the ``1DM'' phase. On the right we display the effects of applying the different bounds to the ``1DM'' phase obtained for a wide scan. For such case there is only one extra mixed scalar so
$|M_{2h}|$ is the mixing matrix element of such non-dark state with the SM Higgs doublet component.
The points have been overlaid following the order in the key (first in the key list is the bottom layer in the plot).}
\end{figure}

Model 2 ($cSM_2$) on the other hand displays more interesting possibilities. Here  we label the phase with one dark matter candidate and a new scalar by ``1DM'', whereas the phase with no dark matter candidates and all three scalars mixed, by ``mix''. 
Fig.~\ref{fig_cSM2F1} (left) shows the phase diagram projection in the plane ($\lambda$,  $|M_{ih}|$), where  $\lambda$ is the doublet quartic coupling and  $|M_{ih}|$  
is the mixing matrix element with the SM Higgs doublet component of (any of) the new mixed scalar(s) which is not the $125$~GeV scalar. Three sets of points are displayed in this
plot (where all bounds have been taken into account): the phase ``mix'' with a wide scan and the phase ``1DM'' both with a standard and a wide scan. The wide and standard scans allow us to
discriminate fixed boundaries from moving boundaries between the two phases. It is clear that the left most boundary is moving towards the vertical axis and therefore cannot be considered a physical boundary between the two phases. This is observed by comparing the boundary between the navy and blue points. Concentrating on the two wide scans (pink and navy colour) it is clearly possible to separate the two phases for some regions of the parameter space; for instance,
for small values of $|M_{ih}|$, $\lambda$ has to be close and above its SM value in phase ``1DM'' while in the ``mix'' phase it can be either above or below that value in a much wider interval. Thus, in this projection, one can clearly identify two pink regions which are exclusive of the ``mix'' phase (excluding the region close to the moving boundary along the vertical axis).

In the right panel of Fig.~\ref{fig_cSM2F1}  we display the effects of applying the different bounds to the ``1DM'' phase obtained for a wide scan. In this case 
the horizontal axis has the variable $|M_{2h}|$ which is the mixing matrix element with the SM Higgs doublet component of the non-dark matter state.
The points have been overlaid following the order in the key - first in the key list is the bottom layer in the plot.
It can be seen that EWPO constraints cut out a considerable portion of the parameters space and are
 responsible for the upper right boundary between the allowed region for the one dark matter phase as compared to the mixed phase in the left panel. 
 Nevertheless, the theoretical constraints alone are responsible for the bottom boundary as seen from comparing the left and 
 the right plots. Note that the apparent top left boundary is not meaningful, and should shrink to the vertical axis if we perform increasingly wider scans in parameter space, as discussed in the previous paragraph.

\begin{figure}
\begin{center}
\includegraphics[clip=true,scale=0.5,trim = 3.8cm 5.2cm 4cm 9.8cm]{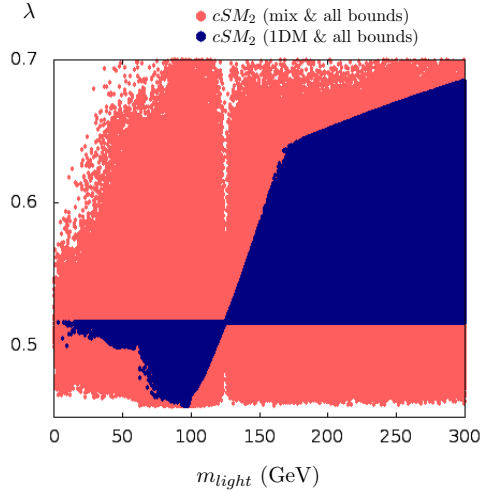}
\end{center}
\caption{\label{fig_cSM2F2}{\em Phase diagram model 2 ($cSM_2$)}: Scatter plot of a projection of the parameter space points obtained for a wide scan of the two phases of model 2. 
$\lambda$ is plotted against $m_{light}$, the lightest non-dark matter state. The points have been overlaid following the order in the key (first in the key list is the bottom layer in the plot).}
\end{figure}
In Fig.~\ref{fig_cSM2F2}  we plot the projection on the  ($\lambda$,  $m_{light}$) plane, where  $m_{light}$ is the mass of  the lightest non-dark matter state.
If some other scalar is detected at the LHC there are values of $\lambda$ that are only allowed in the ``mix'' phase. The SM value for $\lambda$ (fixed
by the Higgs and W-boson masses)  is the horizontal line slightly above 0.5. If the new scalar is lighter than the SM Higgs the ``1DM'' phase only exists below that value while if it heavier than the SM Higgs that phase is only present for values above the SM $\lambda$ . 

\begin{figure}
\begin{center}
\includegraphics[clip=true,scale=0.5,trim = 3.4cm 5.2cm 4cm 8.6cm]{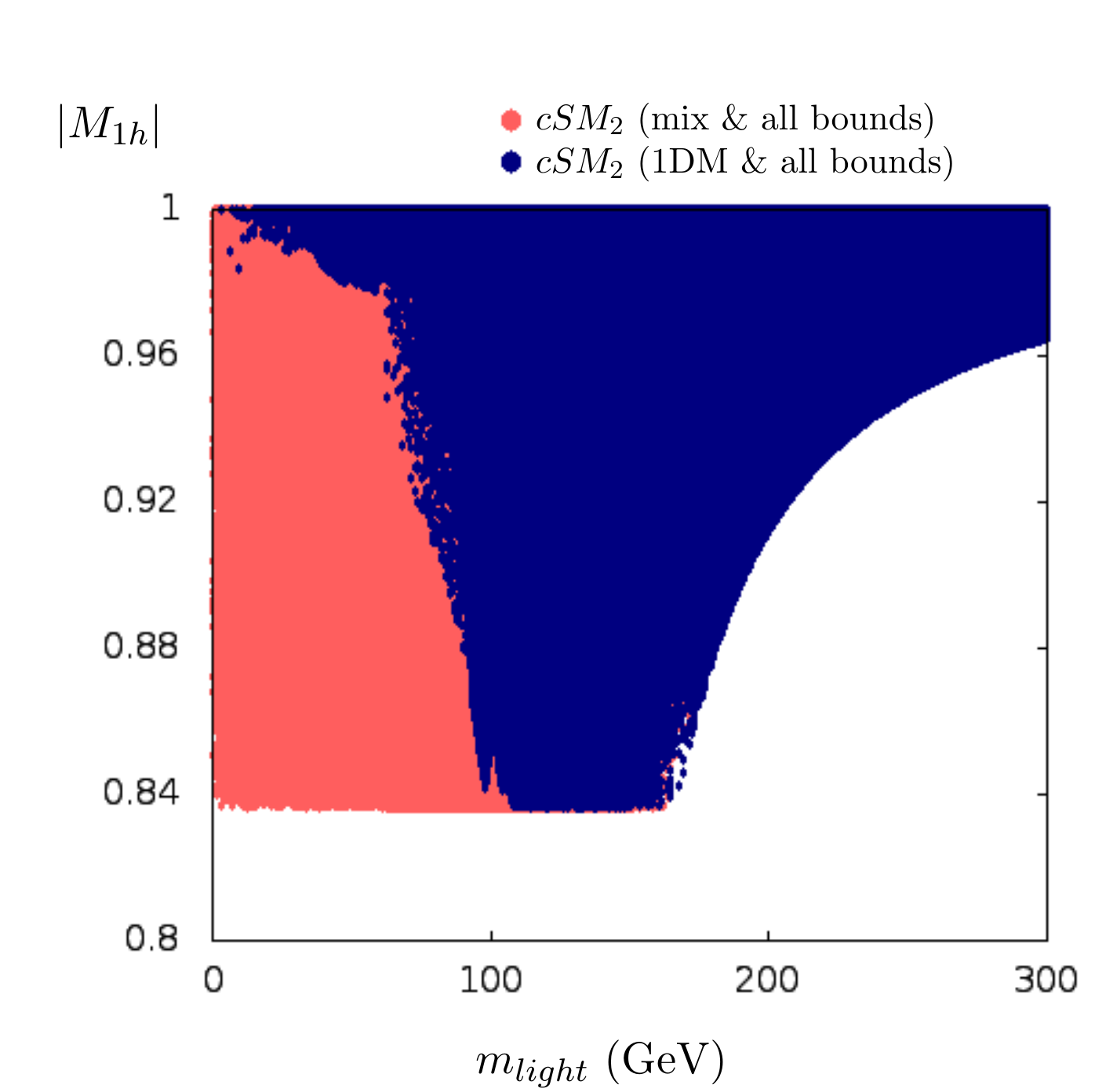}\includegraphics[clip=true,scale=0.5,trim = 2.5cm 5.2cm 4cm 9.2cm]{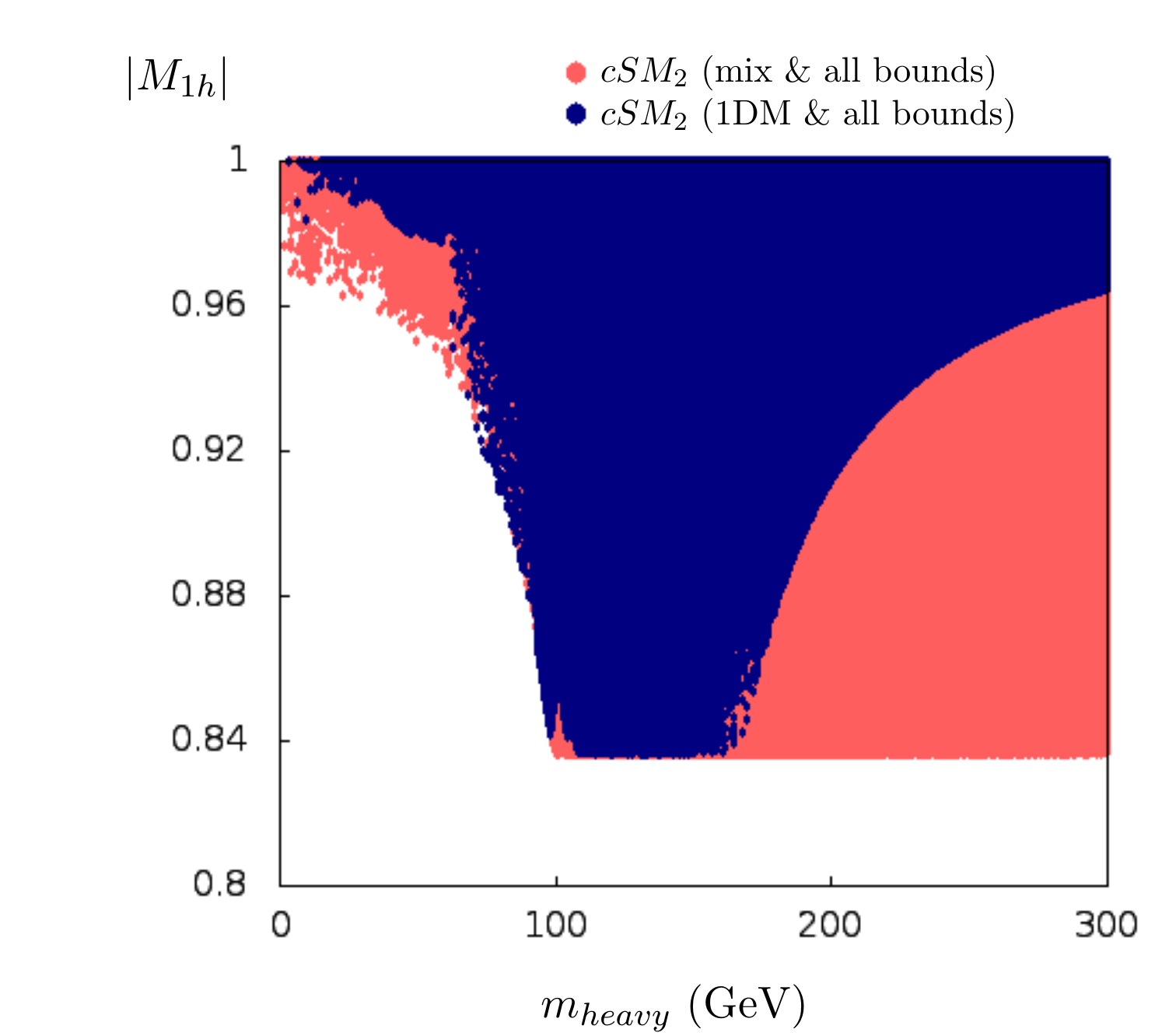} 
\\ \includegraphics[clip=true,scale=0.507,trim = 3.8cm 5.2cm 4cm 8cm]{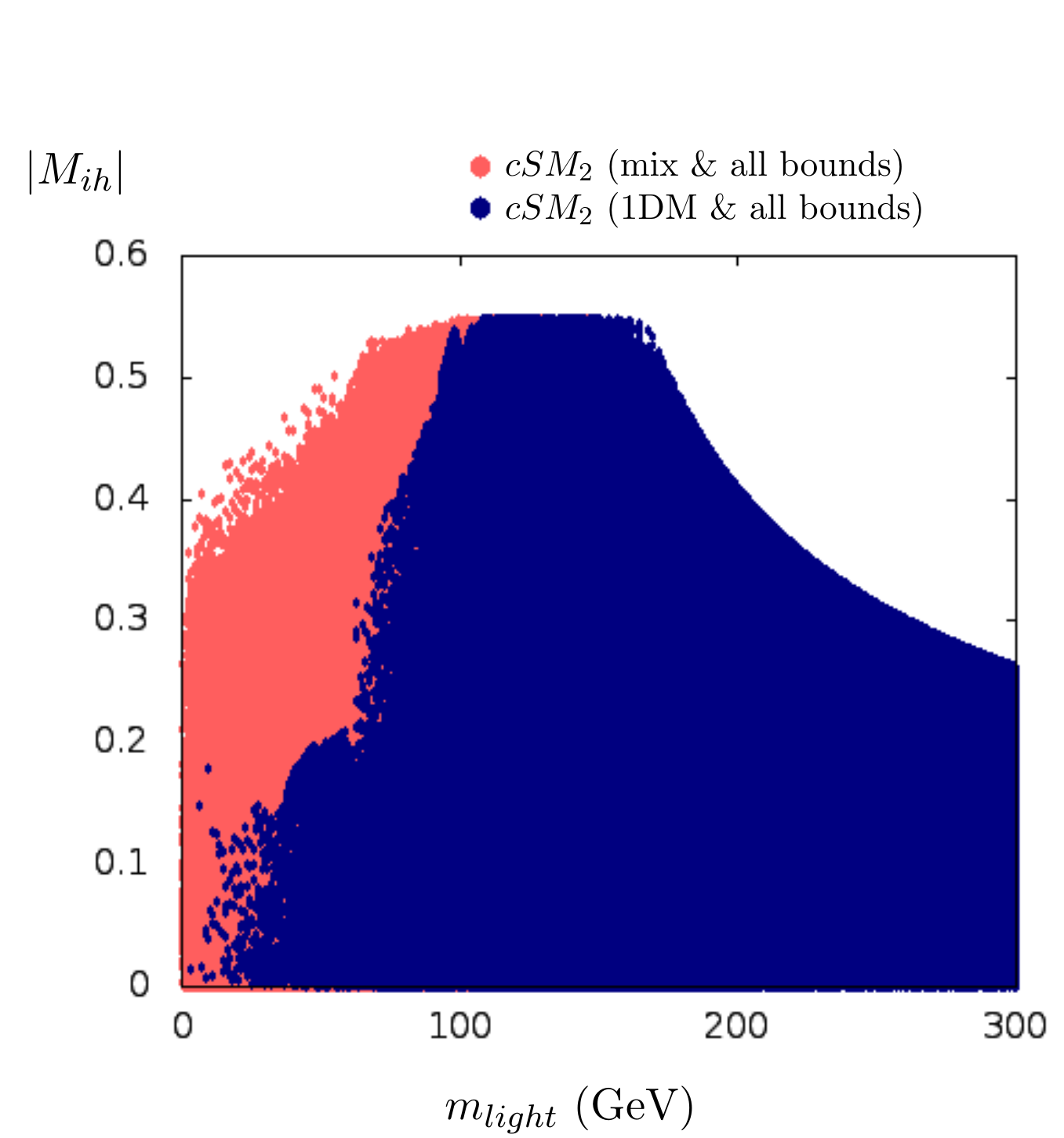}\includegraphics[clip=true,scale=0.507,trim = 2.5cm 5.2cm 4cm 8cm]{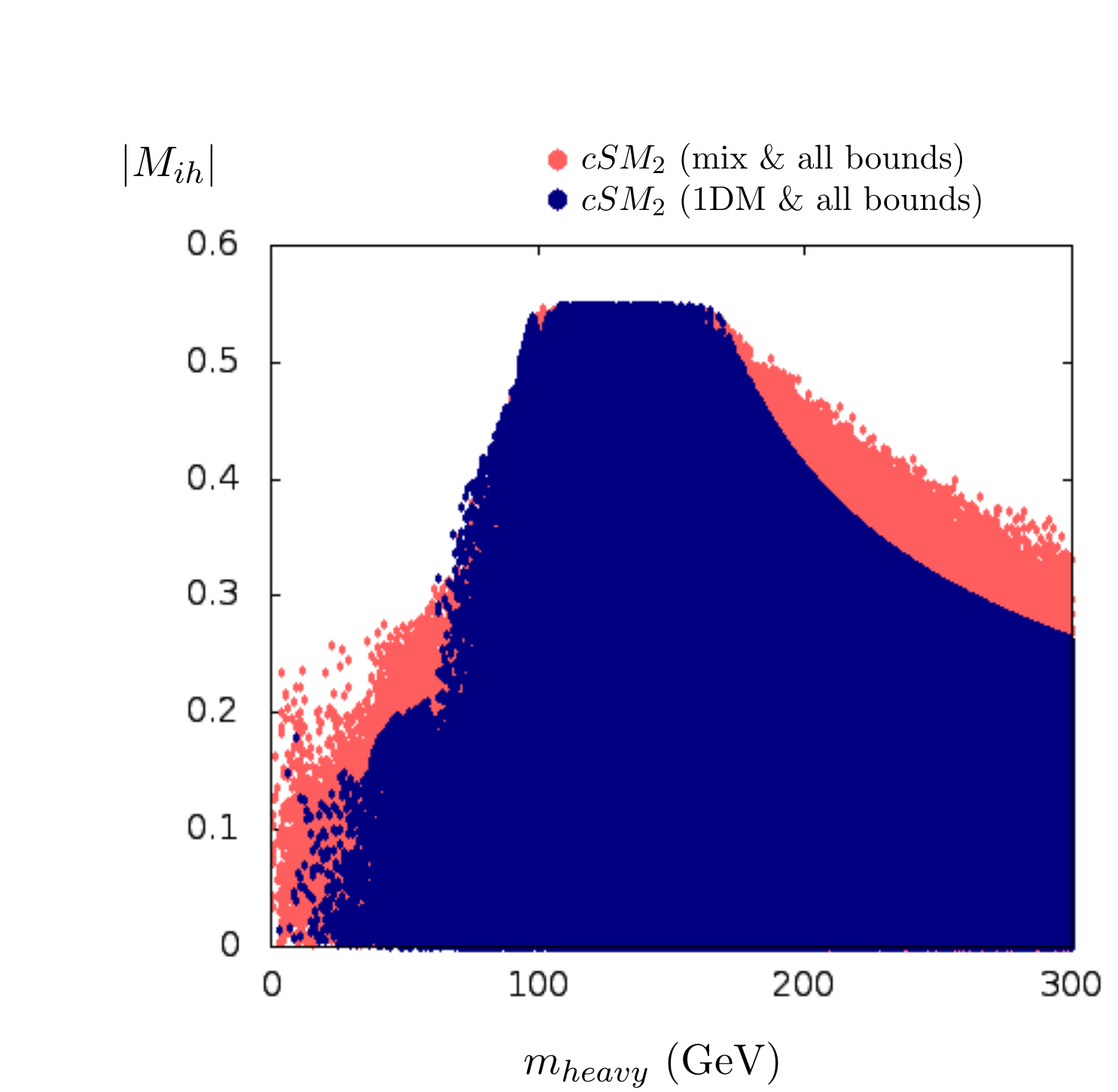} 
\end{center}
\caption{\label{fig_cSM2F3}{\em Phase diagram for model 2 ($cSM_2$)}: Scatter plot of various projections of the parameter space points obtained for a wide scan of the two phases of model 2. 
In the horizontal axis we have either $m_{light}$ or $m_{heavy}$ which are the masses of the lightest or heaviest of the new mixed scalars which is not the $125$~GeV Higgs. 
In the vertical axis we have either $|M_{ih}|$ (mixing matrix element with the SM Higgs doublet component of the non-dark matter scalar corresponding to the mass in the horizontal axis)
or $|M_{1h}|$ which is the 125 GeV scalar mixing matrix element. Note that if the 125 GeV scalar is not allowed to decay to any of the other two scalars, $|M_{1h}|$ is just $\sqrt{\mu}$, that is, 
the square root of the signal strength parameter. The points have been overlaid following the order in the key (first in the key list is the bottom layer in the plot).}
\end{figure}

More interesting are the cases presented in Fig.~\ref{fig_cSM2F3} that show scatter plots 
of various projections of the parameter space points obtained for a wide scan of the two phases of model 2. 
In the horizontal axis we have either $m_{light}$ or $m_{heavy}$ which are the masses of the lightest or heaviest of the new mixed scalars which are neither the $125$~GeV Higgs
nor the dark matter candidate. 
In the vertical axis we have either $|M_{ih}|$ (mixing matrix element with the SM Higgs doublet component of the non-dark matter scalar corresponding to the mass in the horizontal axis),
or $|M_{1h}|$ which is the 125 GeV scalar mixing matrix element. Note that if the 125 GeV scalar is not allowed to decay to any of the other two scalars, $|M_{1h}|$ is just $\sqrt{\mu}$, that is, 
the square root of the signal strength parameter. As before, the points have been overlaid following the order in the key (first in the key list is the bottom layer in the plot).
In this case we are dealing with directly measurable quantities only. Finding a new particle and measuring its decay rates will gives us access to both the masses and $|M_{ih}|$ while
there are already results for $|M_{1h}|$ from the LHC measurements. In fact note that current LHC bound of the signal strength already impose a strong constraint 
on the mixing yielding  $|M_{1h}|\gtrsim 0.84$~\cite{CSearch}. We should note however that this is just the $1\sigma$ bound
for $\sqrt{\mu}$ - had we taken the $2 \sigma$ result the limit on $|M_{1h}|$ would be relaxed. 
  The experimental bounds also force $|M_{ih}|\lesssim 0.55$. The plots in Fig.~\ref{fig_cSM2F3} clearly show 
that a combination of measurements of a new scalar at the LHC may decide for a given phase or exclude the scenario altogether.
The top plots show that a scalar with a mass close the SM Higgs mass is allowed to exist in the ``1DM'' phase. Furthermore, there are regions which are clearly exclusive of 
a no dark matter phase. The same trend appears in the bottom plots even if the distinction between phases is not so striking. A very interesting property, that is observed in both bottom and top plots, is that given a measurement of $|M_{1h}|$ or $|M_{ih}|$ and the mass of the new scalar in a region which is exclusive of the ``mix'' phase, one can infer whether a heavier or lighter scalar is expected to be observed (within this model). For concreteness consider for example the pink (light grey) regions in the left plots of the figure, there is clearly a big portion which does not exist in the corresponding region of the right plots, so if a measurement falls in that region, we can immediately say that if we are in the $cSM_2$ we are looking at the light scalar of the ``mix'' phase. Similarly, there is a pink region in the right plots which do not exist in the left plot, so in such case one can say we are observing the heavy scalar in the ``mix'' phase of the $cSM2$.

\section{Conclusions}

We have presented a new tool, \textsc{ScannerS}, devoted to the 
search for global minima in multi-Higgs models. 
In this work we have applied it to some versions of a simple extension
of the SM - the addition of a complex singlet to the SM doublet, with some symmetries. The code includes the most
relevant theoretical and experimental bounds. Our main focus is in distinguishing the
possible phases of each model by using the present experimental data both from
the LHC and from dark matter experiments.

Once we have identified our working models based on symmetries, we have excluded all phases that did not display the correct electroweak symmetry breaking pattern.  We ended up with two possible phases for each model. In the
first model, which we called model 0, one of the phases leads to a massless dark matter
candidate which is already excluded by the Bullet Cluster results. 
The other phase has two dark matter candidates.
Because in the allowed phase there is no mixing between scalars, the only way
to tell them apart is by actually detecting a dark matter particle.
Hence, a study with \textsc{ScannerS} would add no advantageous information to what we know experimentally.

The cases of model 1 and (especially) of model 2 are the most 
interesting.  We have shown that by measuring physical quantities like
the particle masses, mixing angles, or quartic couplings we are able,
in some particular cases, to pinpoint the phase that is realised in Nature if one of these models applies.
Most importantly, in model 2 a simultaneous measurement of the mass of a  non-dark matter scalar  
together with its mixing angle could be enough to exclude a dark matter phase, and simultaneously indicate whether we are observing the lightest or the heaviest of the new scalar states expected in the model.
Nevertheless, as we move closer and closer to the SM limit the phases become more and more indistinguishable. 

In summary, although the differences that we found between phases of these singlet extensions are restricted only to some measurable quantities, we found it possible to fall into regions where measurements will definitely exclude one of the phases and predict properties of the scalar spectrum.
An interesting question is whether such differences between phases can be identified for more complicated scalar extensions or if they can even become more impressive and predictive.

The ultimate goal of \textsc{ScannerS} is to provide the community with a tool  that can be used 
to search for global minima in general scalar sectors. Although the core routines can already be used for an arbitrary scalar sector, the present release~\cite{ScannerS} requires the user to define the boundedness from below and global minimum routines. In the next release, we expect to provide core routines for such tasks, as well as further analysis examples such as the important case of the two-Higgs doublet model. The present publicly available code includes examples with all the analysis used in this article, which we hope will be a useful starting point for users who intend to explore this tool.

\section*{Acknowledgements}
We would like to thank Augusto Barroso, Pedro Ferreira and Jo\~ao P. Silva for
comments and suggestions.
The work of R.C., M.S. and R.S. is supported in part by the Portuguese
\textit{Funda\c{c}\~{a}o para a Ci\^{e}ncia e a Tecnologia} (FCT)
under contract PTDC/FIS/117951/2010 and by FP7 Reintegration Grant, number PERG08-GA-2010-277025. R.S. is also partially supported by PEst-OE/FIS/UI0618/2011.
 R.C. is funded by FCT through the grant SFRH/BPD/45198/2008.
 M.S. is funded by FCT through the grant SFRH/BPD/ 69971/2010.

\appendix

\section{Stationary point equations}
\label{stationary}
Given a point in parameter space with definite numerical values for 
the couplings $\lambda_a$ and VEVs $\phi_i$,  we need to check for other minima
 which are below the local one we have selected. Solving Eqs.~\eqref{eq:MinCond} for the various cases we obtain
 the following stationary points:
\subsection{Model 1}
\begin{itemize}
\item
 $A=0$, $H=0$, $S^2=-\frac{b_1+b_2}{d_2}$ 
\item
 $\mathbb{S}=0$, in which case we need $H^{\dagger} H=-\frac{m^2}{\lambda}$ 
\item
 $A=0$, $ S^2=\frac{\lambda(b_1+b_2)-\delta_2m^2}{\delta_2^2-d_2\lambda}$, $H^{\dagger} H=-\frac{m^2+\delta_2 S^2}{\lambda}$
\item
 $S=0$, $ A^2=\frac{\lambda(-b_1+b_2)-\delta_2m^2}{\delta_2^2-d_2\lambda}$, $H^{\dagger} H=-\frac{m^2+\delta_2 A^2}{\lambda}$
\end{itemize}
\subsection{Model 2}
\begin{itemize}
\item
  $H=0$, $A=0$ and the following cubic equation must be solved
\begin{equation}
S(b_1+b_2+d_2S^2)+2a_1=0
\end{equation}
\item
  $H=0$, $S=-a_1/b_1$ and 
\begin{equation}
A^2=\frac{b_1^2(b_1-b_2)-d_2a_1^2}{d_2b_1^2}
\end{equation}
\item
  $A=0$, $H^{\dagger} H=-\frac{m^2+\delta_2 S^2}{\lambda}$ and the following cubic equation must be solved
\begin{equation}
S\left[b_1+b_2-\frac{\delta_2m^2}{\lambda}+\left(d_2-\frac{\delta_2^2}{\lambda}\right)S^2\right]+2a_1=0
\end{equation}

\item
  $S=-a_1/b_1$, $H^{\dagger} H=-\frac{m^2+\delta_2 (S^2+A^2)}{\lambda}$ and
\begin{equation}
A^2=\frac{b_1^2(\lambda(b_1-b_2)+m^2\delta_2)-d_2a_1^2\lambda+\delta_2^2a_1^2}{d_2b_1^2\lambda-\delta_2^2b_1^2}
\end{equation}
\end{itemize}

\section{SI scattering cross-section}
\label{SIcs}

\subsection{Model 1 with vanishing singlet VEV}

In this case we have two dark matter candidates, $A$ and $S$ and the couplings with the DM particles and the Higgs state is given by
\begin{equation}
g_{HSS}= g_{HAA} = - \frac{1}{2} \delta_2 v \; .
\end{equation}
The SI scattering cross section is
\begin{equation}
\sigma_{SI} = \frac{m_p^4}{2\pi v^2} \left( \frac{g_{H S S }}{M^2_H}\right)^2 \left( f_{pu} + f_{pd} + f_{ps} + \frac{2}{27} (3 f_G)\right)^2 \left( \frac{1}{(m_p+m_S)^2} +  \frac{1}{(m_p+m_A)^2}\right) \, .
\end{equation}

\subsection{Model 1 with singlet VEV}
In this case we have 1 dark matter candidate, $A$ and two scalars, $H_1$ and $H_2$. The couplings between the DM particle and the scalar eigenstates are
\begin{eqnarray}
g_{AA H_1} &=& (\delta_2 v \cos\phi + d_2 v_S \sin\phi)/2,\\
g_{AA H_2} &=& (d_2 v_S \cos\phi - \delta_2 v \sin\phi)/2
\end{eqnarray}
where $v_S$ is the singlet VEV and $\phi$ the mixing angle.
The expression for $\sigma_{SI}$ is now
\begin{equation}
\sigma_{SI} = \frac{m_p^4}{2\pi v^2 (m_p + m_A)^2} \left( \frac{g_{A A H_1} \cos\phi}{M^2_{H_1}} - \frac{g_{A A H_2} \sin\phi}{M^2_{H_2}} \right)^2 \left( f_{pu} + f_{pd} + f_{ps} + \frac{2}{27} (3 f_G)\right)^2 \, .
\end{equation}

\subsection{Model 2 with $\left<A\right>=0$}

The expression for $\sigma_{SI}$ is the same as in the last subsection.

\section{Ranges for the scans in parameter space}
\label{range}

\begin{table}
\hspace{4.4cm}$\mathbb{Z}_2\times \mathbb{Z}_2'$\hspace{5cm} $\cancel{\mathbb{Z}}_2\times \mathbb{Z}_2'\rightarrow \mathbb{Z}_2'$\\
\begin{tabular}{|c||cc|cc||}
\hline
         & \multicolumn{2}{|c|}{standard run}& \multicolumn{2}{c||}{wide run} \\\hline\hline
coupling & min & max & min & max \\\hline
$m^2$ (GeV$^2$) & $- 10^6$ & $0$ & $-2\,.10^6$ & $0$ \\
$\lambda$ & $0$ & $4$ & $0$ & $50$ \\
$\delta_2$ & $-4$ & $4$ & $-50$ & $50$ \\
$b_2$ (GeV$^2$) & $-10^6$ & $10^6$ & $- 2\,.10^6$ & $2\,.10^6$\\
$d_2$ & $0$ & $4$ & $0$ & $50$ \\
$b_1$ (GeV$^2$)& $-10^6$ & $10^6$ & $- 2\,.10^6$ & $2\,.10^6$\\
$a_1$ (GeV$^3$)& $0$ & $0$ & $0$ & $0$\\ \hline\hline
mass  & min  & max  & min & max \\\hline
$m_h$ (GeV)  & $125$ & $125$ & $125$ & $125$ \\
$m_{D_1}$ (GeV) & $0$ & $300$ & $0$ & $10^3$ \\
$m_{D_2}$ (GeV)& $0$ & $300$ & $0$ & $10^3$ \\ \hline\hline
VEVs  & min & max & min & max \\\hline
$v$ (GeV) & $246$ & $246$ & $246$ & $246$ \\
$v_S$ (GeV) & $0$ & $0$ & $0$ & $0$ \\
$v_A$ (GeV) & $0$ & $0$ & $0$ & $0$ \\ \hline
\end{tabular}
\begin{tabular}{||c|cc|cc|}
\hline
  &    \multicolumn{2}{|c|}{standard run}& \multicolumn{2}{c|}{wide run} \\\hline\hline
 &min & max & min & max \\\hline
  &$- 10^6$ & $10^6$ & $-2\,.10^6$ & $2\,.10^6$ \\
  &$0$ & $4$ & $0$ & $50$ \\
  &$-4$ & $4$ & $-50$ & $50$ \\
  &$-10^6$ & $10^6$ & $- 2\,.10^6$ & $2\,.10^6$\\
  &$0$ & $4$ & $0$ & $50$ \\
  &$-10^6$ & $10^6$ & $- 2\,.10^6$ & $2\,.10^6$\\
  &$0$ & $0$ & $0$ & $0$\\ \hline\hline
  &min  & max  & min & max \\\hline
$m_{h}$ &$125$ & $125$ & $125$ & $125$ \\
 $m_{H_1}$ &$0$ & $300$ & $0$ & $10^3$ \\
  $m_{D}$&$0$ & $300$ & $0$ & $10^3$ \\ \hline\hline
  &min & max & min & max \\\hline
  &$246$ & $246$ & $246$ & $246$ \\
  &$0$ & $500$ & $0$ & $10^3$ \\
  &$0$ & $0$ & $0$ & $0$ \\ \hline
\end{tabular} \\
\caption{Parameter ranges for $cSM_1$ {\em Left:} 2DM symmetric phase , {\em Right:} 1DM broken phase.}
\label{tab:range1}
\end{table}

In this appendix we present the ranges that were used for the scans over the couplings $\lambda_a$, masses and VEVs of the scalars.  We have performed two 
different scans, which are indicated in the tables respectively by ``standard run'' and ``wide run''. In the latter we allow the hypercubic box in parameter space to be wider to investigate which boundaries in the phase diagram do not changed significantly. Each table contains the two possible phases of each model indicated appropriately. 

Table~\ref{tab:range1}, {\em Left}, shows the ranges of couplings, masses and VEVs of the scalars in the symmetric phase 
``2DM'' of model $cSM_1$ (two dark matter phase) and for the two different scans (standard and wide). $h$ 
refers to the Standard Model Higgs and $D_1$ and $D_2$ the two dark matter candidates of the model. Table~\ref{tab:range1}, {\em Right}, is for the same model ($cSM_1$) but in the broken phase where there is only one dark matter particle (1DM); 
$h$ is the Standard Model Higgs which mixes with the scalar $H_1$ and $D$ the dark matter candidate.
\begin{table}
\hspace{4.9cm}$\mathbb{Z}_2'$ \hspace{6.35cm}  $\cancel{\mathbb{Z}}_2'$ \\
\begin{tabular}{|c||cc|cc||}
\hline
& \multicolumn{2}{c|}{standard run}& \multicolumn{2}{c||}{wide run} \\\hline\hline
coupling & min & max & min & max \\\hline
$m^2$ (GeV$^2$) & $- 10^6$ & $10^6$ & $-2\,.10^6$ & $2\,.10^6$ \\
$\lambda$ & $0$ & $4$ & $0$ & $50$ \\
$\delta_2$ & $-4$ & $4$ & $-50$ & $50$ \\
$b_2$ (GeV$^2$) & $-10^6$ & $10^6$ & $- 2\,.10^6$ & $2\,.10^6$\\
$d_2$ & $0$ & $4$ & $0$ & $50$ \\
$b_1$ (GeV$^2$)& $-10^6$ & $10^6$ & $- 2\,.10^6$ & $2\,.10^6$\\
$a_1$ (GeV$^3$)& $-10^6$ & $10^6$ & $- 10^8$ & $10^8$\\ \hline\hline
mass  & min  & max  & min & max \\\hline
$m_h$ (GeV) & $125$ & $125$ & $125$ & $125$ \\
$m_{H_1}$ (GeV) & $0$ & $300$ & $0$ & $10^3$ \\
$m_{D}$ (GeV) & $0$ & $300$ & $0$ & $10^3$ \\ \hline\hline
VEVs  & min & max & min & max \\\hline
$v$ (GeV) & $246$ & $246$ & $246$ & $246$ \\
$v_S$ (GeV) & $0$ & $500$ & $0$ & $10^3$ \\
$v_A$ (GeV) & $0$ & $0$ & $0$ & $0$ \\ \hline
\end{tabular}
\begin{tabular}{||c|cc|cc|}
\hline
      &\multicolumn{2}{||c|}{standard run}& \multicolumn{2}{c|}{wide run} \\\hline\hline
 &min & max & min & max \\\hline
 &$- 10^6$ & $10^6$ & $-2 \, .10^6$ & $2\,.10^6$ \\
 &$0$ & $4$ & $0$ & $50$ \\
 &$-4$ & $4$ & $-50$ & $50$ \\
 &$-10^6$ & $10^6$ & $- 2\,.10^6$ & $2\,.10^6$\\
 &$0$ & $4$ & $0$ & $50$ \\
 &$-10^6$ & $10^6$ & $- 2\,.10^6$ & $2\,.10^6$\\
 &$-10^6$ & $10^6$ & $- 10^8$ & $10^8$\\ \hline\hline
 &min  & max  & min & max \\\hline
 $m_h$ &$125$ & $125$ & $125$ & $125$ \\
 $m_{H_1}$&$0$ & $300$ & $0$ & $10^3$ \\
 $m_{H_2}$&$0$ & $300$ & $0$ & $10^3$ \\ \hline\hline
 &min & max & min & max \\\hline
 &$246$ & $246$ & $246$ & $246$ \\
 &$0$ & $500$ & $0$ & $10^3$ \\
 &$0$ & $500$ & $0$ & $10^3$ \\ \hline
\end{tabular}\\
\caption{Parameter ranges for $cSM_2$. {\em Left:} ``1DM'' symmetric phase, {\em Right:} ``mix'' broken phase.}
\label{tab:range2}
\end{table}

The last table is for model 2 ($cSM_2$). Table~\ref{tab:range2}, {\em Left}, is for the symmetric phase, which contains a dark matter particle (1DM). Similarly to model 1, the scalar spectrum is given by the Standard Model Higgs, $h$, which mixes with the scalar $H_1$ and $D$ is the DM candidate. For the broken phase, Table~\ref{tab:range2}, {\em Right}, the scalar spectrum contains the Standard model Higgs, $h$,  which mixes with the two scalars $H_1$ and $H_2$.

\bibliography{references}
\bibliographystyle{JHEP}

\end{document}